\documentclass[11pt,letterpaper]{article}
\usepackage[letterpaper,dvips,body={8in,11in},vmargin={1cm,2cm},hmargin=2cm,head=1cm]{geometry}
\usepackage[dvips,usenames]{color}
\usepackage[american]{babel}
\usepackage[latin1]{inputenc}
\usepackage[T1]{fontenc}
\usepackage{ae}
\usepackage{aecompl}
\usepackage{textcomp} 
\usepackage{textfit}  
\usepackage{xspace}
\usepackage{graphicx}
\usepackage{latexsym}
\usepackage{amsmath,amsfonts,amstext,amssymb,amsbsy,amsopn,amsthm,eucal}
\usepackage{yfonts}[1998/10/03]
\usepackage{dsfont}
\usepackage{wasysym}
\usepackage{wrapfig}
\usepackage[normalem]{ulem}
\usepackage{url}
\newcommand\email{\begingroup \urlstyle{tt}\Url}
\urldef{\het}\url{www.het.brown.edu}
\urldef{\danieldf}{\email}{danieldf@het.brown.edu}
\newcommand{\cont}[1]{\ensuremath{\text{\rsfs C\/}^{#1}\xspace}}

\newcommand{\Matrix}[2]{\ensuremath{\text{\rsfs M}_{#1}(\mathbb{#2})}\xspace}
\newcommand{\Cliff}[3]{\ensuremath{\text{\rsfs C}_{#1,#2}(\mathbb{#3})}\xspace}
\newcommand{\ctor}{\ensuremath{\text{\rsfs C}_{3,1}(\mathbb{R})}\xspace}

\newcommand{\mfr}{\ensuremath{\text{\rsfs M}_{4}(\mathbb{R})}\xspace}

\newcommand{\Alg}[1]{\ensuremath{\text{\rsfs A}\xspace(\mathds{#1})\xspace}}
\newcommand{\Lin}[2]{\ensuremath{\text{\rsfs L}\xspace(\mathds{#1},\mathds{#2})\xspace}}

\DeclareMathOperator{\diag}{diag}

\DeclareMathOperator{\dop}{d\!}

\DeclareMathOperator{\End}{End}

\newcommand{\accentovetor}{%
  \makebox[0cm][c]{%
  \raisebox{-0.53\baselineskip}{\ensuremath{\vec{}}}}}
\newcommand{\Bvec}[1]{\ensuremath{#1\makebox[0cm][r]{\accentovetor\ }}}

\newtheorem{thm}{Theorem}[section]

\newtheorem*{conjec}{Conjecture}
\newtheorem{deff}[thm]{Definition}
\newtheorem*{notat}{Notation}

\newcommand{\paslash}{\ensuremath \raisebox{0.025cm}{\slash}\hspace{-0.25cm}\partial\/}
\newcommand{\dfrakslash}{\ensuremath \slash\hspace{-0.20cm} \mathfrak{d}\/}
\newcommand{\Dslash}{\ensuremath \raisebox{0.025cm}{\slash}\hspace{-0.32cm} D}
\newcommand{\dslash}{\not{\hbox{\kern-2pt $\partial$}}}
\newcommand{\pslash}{\not{\hbox{\kern-2.3pt $p$}}}
 \newtoks\nslashfraction
 \nslashfraction={.13}
 \newcommand{\nslash}[1]{\setbox0\hbox{$ #1 $}
   \setbox0\hbox to \the\nslashfraction\wd0{\hss \box0}/\box0 }

\newcommand{\plpl}{\raise-2pt\hbox{$\raise3pt\hbox{$_+$}\hskip-6.67pt\raise0.0pt
  \hbox{$^+$}\hskip 0.01pt$}}
\newcommand{\mimi}{\raise-2pt\hbox{$\raise3pt\hbox{$_-$}\hskip-6.67pt\raise0.0pt
  \hbox{$^-$}\hskip 0.01pt$}}

\newcommand{\bo}{\raise-1mm\hbox{\Large$\Box$}}              
\newcommand{\pa}{\partial}                                       
\newcommand{\de}{\nabla}                                         
\newcommand{\dg}{\sp\dagger}                                     
\newcommand{\trans}[1]{{#1}^{\ensuremath{\mathsf{T}}}}           
\newcommand{\hc}[1]{{#1}^{\dg}}                            
\newcommand{\nTH}{{\raise.2ex\hbox{$\displaystyle \bigodot$}\mskip-4.7mu \llap H \;}}
\newcommand{\face}{{\raise.2ex\hbox{$\displaystyle \bigodot$}\mskip-2.2mu \llap {$\ddot
        \smile$}}}                                      

\newcommand{\ket}[1]{\left| #1\right\rangle}              
\newcommand{\ev}[1]{\left\langle #1\right\rangle}        
\newcommand{\ip}[2]{\left\langle #1 | #2\right\rangle}    
\newcommand{\ipop}[3]{\left\langle#1\left|#2\right|#3\right\rangle} 
\newcommand{\abs}[1]{\left| #1\right|}                    
\newcommand{\norm}[1]{\left\lVert#1\right\rVert}
\newcommand{\comm}[2]{\left[\,#1,#2\,\right]}                 
\newcommand{\acomm}[2]{\left\{\,#1,#2\,\right\}}              
\newcommand{\leftrightarrowfill}{$\mathsurround=0pt \mathord\leftarrow \mkern-6mu
        \cleaders\hbox{$\mkern-2mu \mathord- \mkern-2mu$}\hfill
        \mkern-6mu \mathord\rightarrow$}
\newcommand{\dvec}[1]{\vbox{\ialign{##\crcr
        \leftrightarrowfill\crcr\noalign{\kern-1pt\nointerlineskip}
        $\hfil\displaystyle{#1}\hfil$\crcr}}}           

\newcommand{\sfrac}[2]{{\vphantom1\smash{\lower.5ex\hbox{\small$#1$}}\over
        \vphantom1\smash{\raise.4ex\hbox{\small$#2$}}}} 
\newcommand{\bfrac}[2]{{\vphantom1\smash{\lower.5ex\hbox{$#1$}}\over
        \vphantom1\smash{\raise.3ex\hbox{$#2$}}}}       
\newcommand{\afrac}[2]{{\vphantom1\smash{\lower.5ex\hbox{$#1$}}\over#2}}    

\newskip\humongous \humongous=0pt plus 1000pt minus 1000pt

\newif\ifdtup

\newfont{\go}{ygoth.tfm scaled 1200}                   
\newfont{\biggo}{ygoth.tfm scaled 3583}                
\newfont{\rope}{cmsy10 scaled 1200}                    
\newfont{\fib}{cmfi10 scaled 1200}
\newfont{\bigfib}{cmfi10 scaled 3583}
\newfont{\funny}{cmff10 scaled 1200}
\newfont{\bigfunny}{cmff10 scaled 3583}
\newfont{\pbk}{pbkd.tfm scaled 1200}

\newfont{\rsfs}{rsfs10.tfm scaled 1200}
\newfont{\bigrsfs}{rsfs10.tfm scaled 2000}

\newfont{\testea}{cmfrak.tfm}
\newfont{\testeb}{dcfrak.tfm}
\newfont{\testec}{schwell.tfm}
\newfont{\tested}{yfrak.tfm}
\newfont{\testef}{yswab.tfm}
\newfont{\bigtestef}{yswab.tfm scaled 3583}
\newfont{\testeg}{yinit.tfm}
\newfont{\testeh}{yinitdd.tfm}
\newfont{\testei}{suet14.tfm}
\newfont{\testej}{pzdr.tfm}
\newfont{\testek}{pzcmi.tfm}
\newfont{\testem}{ccr10.tfm}
\newfont{\testen}{eurm10.tfm}
\newfont{\testeq}{euex10.tfm}
\newfont{\testeo}{wncyr10.tfm}
\newfont{\testep}{msam10.tfm}

\catcode`@=11
\newcommand{\un}[1]{\relax\ifmmode\@@underline#1\else
        $\@@underline{\hbox{#1}}$\relax\fi}
\catcode`@=12

\usepackage{fancyhdr}
\fancypagestyle{plain}{%
\fancyhf{} 
\fancyfoot[C]{\bfseries\thepage} 

}

\newcommand{\helv}{%
\fontfamily{phv}\fontseries{b}\fontsize{9}{11}\selectfont}
\begin{document}
\begin{titlepage}
  \begin{flushright}
    {BROWN-HET-1374 \\ July 2003} \\
    \texttt{math-ph/0308004}
  \end{flushright}
  \bigskip
  \begin{center}
    \huge{Generalized Bundle Quantum Mechanics} \\
    \vspace{1cm}
    \Large{Daniel D. Ferrante\footnote{E-mail: \danieldf}} \\
    \medskip
    HET --- Physics Department, Brown University, USA\footnote{URL: \hspace*{0.3cm}\het}.  
  \end{center}
  \vspace{2cm}
  \tableofcontents
\end{titlepage}
\nocite{fbfqm01, fbfrqm02, gp01, gcroaqtcs01, cagc84, isg89, gcfbqm}
\setcounter{footnote}{0}
\section{Generalized Bundle Quantum Mechanics}\label{ch:gbqm}
\subsection{Mathematical Background}\label{sec:mb}
In this section, a quick review of the mathematical background shall be presented to the
reader. The material includes:

\begin{itemize}
\item Vector spaces, duals and double duals;
\item Tensors over a vector space;
\item Exterior algebras $\Lambda^{*}$ and $\Lambda$;
\item Metric and tensors;
\item Clifford algebras and spinors;
\item Fibre bundles as special manifolds;
\item Geometrical definition of \ctor spinors;
\item Differentiation of spinor fields;
\end{itemize}

\noindent The reader that is already familiar with it, can just jump ahead to section
\ref{sec:gcopqm}.
\subsection{Vector spaces, duals and double duals}\label{subsec:vsddd}
A vector space can be defined as,

\begin{deff}
  A set $\mathds{V}$ is called a \emph{vector [linear] space}, $(V,+,\cdot,\mathbb{K})$, over
  a [scalar] field $\mathbb{K}$ if, between two elements of $\mathds{V}$, there is an
  \emph{addition}, $+$, defined and, between and element of $\mathds{V}$ and an element of
  $\mathbb{K}$, there is a \emph{scalar multiplication}, $\cdot\;$. In addition to these,
  $\mathds{V}$ should be \emph{closed} under addition and scalar multiplication. (Not to
  mention the other 4 properties of addition and the other 4 properties of scalar
  multiplication: associativity, distributivity, commutativity and the neutral element.)
\end{deff}

On top of the definition of a vector space, an algebra can be defined. Here's how one does
it:
\begin{deff}
  An \emph{algebra}, $\Alg{V} = (\mathds{V},+,\cdot,\odot,\mathbb{K})$, is a mathematical object
  constructed out of a vector space by endowing it with an \emph{internal} multiplication
  rule, i.e., an operation \; $\odot: V \times V \to V$ that satisfies:
  \begin{align*}
    \vec{v}\odot(a\,\vec{x} + b\,\vec{y}) &= a\,(\vec{v}\odot\vec{x}) + b\,(\vec{v}\odot\vec{y}) \\
    (a\,\vec{x} + b\,\vec{y})\odot\vec{v} &= a\,(\vec{x}\odot\vec{v}) + b\,(\vec{y}\odot\vec{v})\, ,
      \quad \forall \; \vec{v},\vec{x},\vec{y}\in \mathds{V}\; \text{and}\;\; \forall \; a,b\in \mathbb{K}
      \; .\\
  \end{align*}
\end{deff}

Now, a vector space can have a \emph{basis}. For that, it suffices to find a set of $n$
\emph{linearly independent} vectors, $\{\vec{e}_i,\; 1\leqslant i \leqslant n\}$, such that
$\forall \; \vec{v}\in\mathds{V}$ can be uniquely written as $\vec{v} = \sum_{i=1}^{n}v^i\,
\vec{e}_i, \; v^i\in\mathbb{K}$. In this case, the $\{v^i\}$ are called the \emph{components}
of $\vec{v}$ in the basis $\{\vec{e}_{(i)}\}$ and $n = \text{dim}(\mathds{V})$. (Our focus
will be on finite $n$ vector spaces.)

\begin{notat}\label{notat:1}
  For convenience sakes, the following will be used:

  \begin{align*}
    \vec{v} &= \sum_{i=1}^{n}v^i\, \vec{e}_i = \vec{e}_{\bullet}\, v^{\bullet} \\
    \intertext{where}
    \vec{e}_{\bullet} &\equiv (\vec{e}_1,\dotsc,\vec{e}_n) \\
    v^{\bullet} &\equiv
    \begin{pmatrix}
      v^1 \\
      \vdots \\
      v^n
    \end{pmatrix}
  \end{align*}
\end{notat}

The concept of a \emph{linear mapping} [between two vector/linear spaces] is a quite
useful one, so let's get to it.

\begin{deff}
  A \emph{linear mapping} between $(\mathds{V},+,\cdot,\mathbb{K})$ and
  $(\mathds{W},+,\cdot,\mathbb{K})$ is a mapping that satisfies:

  \begin{align*}
    \mu:\quad\quad\; \mathds{V} &\to \mathds{W} \\
    \vec{v}_1 + \vec{v}_2 &\mapsto \mu(\vec{v}_1) + \mu(\vec{v}_2) \\
    a\, \vec{v} &\mapsto a\, \mu(\vec{v})
  \end{align*}
\end{deff}

Using this, the set of linear mappings from $\mathds{V}$ to $\mathds{W}$, denoted by
\Lin{V}{W}, is a vector space (addition is given by the addition of the mappings and
scalar multiplication is given by multiplying the map by the scalar). The space \Lin{V}{V}
is an algebra (\emph{Why?}), where the internal multiplication is given by function
composition.

Thus, in this fashion, $\forall\; \vec{v}\in\mathds{V},\; \vec{v}=\vec{e}_{\bullet}\, v^{\bullet}$, the image
vector is given by:

\begin{align*}
  \vec{\omega} &= \mu(\vec v) = \vec{f}_{\bullet}\, \mu^{\bullet}_{\bullet}\, v^{\bullet} \\
  \intertext{or, in matrix language,}
  \omega^{\bullet} &= \mu^{\bullet}_{\bullet}\, v^{\bullet} \\
\end{align*}

\begin{deff}
  The space of the \emph{linear forms} defined on $\mathds{V}$ is called the \emph{dual}
  of $\mathds{V}$, denoted by $\mathds{V}^{*}$. A \emph{linear form} is a linear mapping
  that satisfies:

  \begin{align*}
    b\, \Bvec{v} + c\, \Bvec{\omega}:\qquad \mathds{V} &\to \mathbb{K} \\
    \vec{x} + \vec{y} &\mapsto b\, \Bvec{v}[\vec{x}] + b\, \Bvec{v}[\vec{y}] + c\,
      \Bvec{\omega}[\vec{x}] + c\, \Bvec{\omega}[\vec{y}] \\
    a\, \vec{x} &\mapsto a\,b\, \Bvec{v}[\vec{x} + \vec{y}] + a\, c\, \Bvec{\omega}[\vec{x} +
      \vec{y}]
  \end{align*}
\end{deff}

The reader should note that $\mathds{V}^{*} = \text{\rsfs
  L}\xspace(\mathds{V},\mathbb{K})$.

A basis can, now, be constructed for $\mathds{V}^{*}$ with the help of $\{\vec{e}_i\}$, the
basis for $\mathds{V}$. Thus, $\{\Bvec{e}^i\}$ will be a basis for $\mathds{V}^{*}$, defined
by:

\begin{align*}
  \Bvec{e}^i[\vec{e}_j] &= \delta^i_j \\
  &=
  \begin{cases}
    1, & \text{if}\;\; i = j \\
    0, & \text{if}\;\; i \neq j
  \end{cases}
\end{align*}

In this way, now one can find oneself in the position to calculate the scalar associated
by $\Bvec{v}$ to $\vec{x}$:

\begin{equation*}
  \Bvec{v}[\vec{x}] = v_i\, \Bvec{e}^i[x^j\, \vec{e}_j] = v_i\, x^j\, \delta^i_j = v_i\, x^j \;.
\end{equation*}

In an analogous manner done before, one can define the following notation.

\begin{notat}
  \begin{align*}
    \Bvec{v} &= v_{\bullet}\, \Bvec{e}^{\bullet} \\
    \Bvec{v}[\vec{x}] &= v_{\bullet}\, \Bvec{e}^{\bullet} [\vec{e}_{\bullet}\, x^{\bullet}] = v_{\bullet}\,
      (\Bvec{e}^{\bullet} \,\vec{e}_{\bullet})\, x^{\bullet} \\
    &= v_{\bullet}\,\mathds{1}\, x^{\bullet} = v_{\bullet}\, x^{\bullet} \; ,
    \intertext{where}
    \mathds{1} &= \Bvec{e}^{\bullet} \,\vec{e}_{\bullet} = \text{unit matrix} \; .
  \end{align*}

  The nomenclature of \emph{contravariant} vectors [for elements of $\mathds{V}$] and
  \emph{covariant} vectors [for elements of $\mathds{V}^*$] follows from the fact that,
  upon a change of basis of $\mathds{V}$ by a matrix $A$, the components of the vectors
  transform with $A^{-1}$, while the components of the linear forms transform with $A$.
\end{notat}

With all this material, it is possible, now, to define the \emph{double dual} of
$\mathds{V}$, denoted by $\mathds{V}^{**}$. This will be given by,

\begin{align*}
  \mathbf{v}_{\vec{v}}: \quad \mathds{V}^{*} &\to \mathds{K} \\
  \Bvec{x} &\mapsto \mathbf{v}_{\vec v}[\Bvec{x}] \equiv \Bvec{x}[\vec{v}]
\end{align*}

Therefore, it is possible to prove that $\mathds{V}$ is \emph{isomorphic} to
$\mathds{V}^{**}$!
\subsection{Tensors over a vector space}\label{subsec:tvs}
Once the concept of vectors and linear forms is settled, the next step is going towards
the definition of tensors, which generalize the former. Tensors and spinors will play a
fundamental role in the remainder of this work.

First off, let's start by defining the space of tensors, denoted by $\Theta^p_q$.

\begin{deff}
  A \emph{tensor} ${}^{p}_{q}T$, $p$ times \emph{contravariant} and $q$ times
  \emph{covariant}, also termed a $(p,q)$\emph{-tensor} or a \emph{tensor of rank}
  $(p,q)$, is defined as:

  \begin{align*}
    {}^{p}_{q}T: \quad \underbrace{(\mathds{V}^*\times\dotsb \times\mathds{V}^*)}_{\text{p
      factors}}\times\underbrace{(\mathds{V}\times\dotsb \times\mathds{V})}_{\text{q factors}} &\to
      \mathbb{K} \\
    \big(\Bvec{v}{}_{1},\dotsc,\Bvec{v}{}_p;\vec{\omega}{}_{1},\dotsc,\vec{\omega}{}_q\big) &\mapsto
      {}^{p}_{q}T\big(\Bvec{v}{}_1,\dotsc,\Bvec{v}{}_p;\vec{\omega}{}_{1},\dotsc,\vec{\omega}{}_q\big)
  \end{align*}
  which is \emph{linear} in each entry.
\end{deff}

\begin{notat}
  The set of $(p,q)$-tensors will be denoted by,

  \begin{equation*}
    \Theta^p_q \equiv \big(\mathds{V}^{\otimes p}\big)\otimes\big(\mathds{V}^*\big)^{\otimes q} \; .
  \end{equation*}
\end{notat}

It is not difficult to see that $\Theta^p_q$ is a vector space, $(\Theta^p_q,+,\cdot)$ whose dimension
is given by: $\dim{\mathds{V}} = n\, \Rightarrow\, \dim{\Theta^p_q}=n^{p+q}$.

\begin{deff}
  The \emph{tensor product} (distributive and associative), $\otimes$, is defined as follows:

  \begin{align*}
    \otimes: \quad \Theta^{p_{1}}_{q_{1}}\times\Theta^{p_{2}}_{q_{2}} &\to \Theta^{p_{1}+p_{2}}_{q_{1}+q_{2}} \\
    \big({}^{p_1}_{q_1}S;{}^{p_2}_{q_2}T\big) &\mapsto {}^{p_1}_{q_1}S \otimes {}^{p_2}_{q_2}T
  \end{align*}
\end{deff}

\begin{notat}
  As has been made clear by the notation employed previously, a $(1,1)$-tensor can be
  written as $\Bvec{\vec T}$ (and so on and so forth). A typical example of a
  $(1,1)$-tensor is a linear mapping from a vector space to itself.
\end{notat}

The \emph{contraction} of tensors can also be constructed.

\begin{deff}
  The \emph{contraction mapping}, $C^r_s$, applied on a tensor ${}^p_qT\in\Theta^p_q$, (where
  $1\leqslant r\leqslant p$ and $1\leqslant s\leqslant q$), reduces the contravariant and
  covariant indices by one, thus

  \begin{align*}
    C^r_s:\quad \Theta^p_q &\to \Theta^{p-1}_{q-1} \\
    {}^p_qT = \vec{v}_1\otimes\dotsb\otimes\vec{v}_p\otimes\Bvec{\omega}{}_1\otimes\dotsb\otimes\Bvec{\omega}{}_q &\mapsto
      C^r_s\big({}^p_qT\big) =
      \Bvec{\omega}{}_{s}[\vec{v}_r]\cdot\vec{v}_1\otimes\dotsb\otimes\vec{v}_{r-1}\otimes\vec{v}_{r+1}
      \otimes\dotsb\otimes\vec{v}_p\otimes \\
      &\hspace{3.5cm} \otimes\Bvec{\omega}{}_{1}\otimes\dotsb\otimes\Bvec{\omega}{}_{s-1}\otimes\Bvec{\omega}{}_{s+1}\otimes\dotsb\otimes\Bvec{\omega}{}_{q}
  \end{align*}
\end{deff}
\subsection{Exterior algebras $\Lambda^{*}$ and $\Lambda$}\label{subsec:ea}
Now, the attention will be restricted to special kinds of tensors, the $q$\emph{-linear
  alternating forms}. The \emph{exterior product} will be defined and, thus, the
\emph{Exterior Algebra}, $(\Lambda^*,+,\cdot,\land)$, will be defined.

\begin{deff}
  An \emph{alternating $q$-form} is defined over a [q-linear form] ${}_q\alpha\in\Theta^0_q$ when,

  \begin{equation*}
    {}_q\alpha[\vec{v}_{\sigma(1)},\dotsc,\vec{v}_{\sigma(q)}] = (-1)^{|\sigma|}\,
    {}_q\alpha[\vec{v}_{1},\dotsc,\vec{v}_{q}] \; , \; \forall \vec{v}_i\in\mathds{V}, \; 1\leqslant i
      \leqslant q, \; \forall \sigma\in\mathcal{S}_q, \; q\geqslant 1 \; ,
  \end{equation*}
  where $\mathcal{S}_q$ is the permutation group of the elements $\{1,\dotsc,q\}$, $\sigma$ is
  one particular permutation and $|\sigma|$ is its parity.
\end{deff}

Now, $\Lambda^*_q$ is the subset of $\Theta^0_q$ that comprises those q-linear alternating forms. In
particular, $\Lambda^*_1 \simeq V^*$ (for convenience, the elements ${}_1\alpha\in\Lambda^*_1$ will be denoted as
$\Bvec{\alpha}$) and $\Lambda^*_0 \simeq \mathbb{K}$.

The next step towards the construction of an algebra is defining an internal
multiplication, that will be called \emph{exterior product}. This can be done with the aid
of the tensor product,$(\otimes)$, defined above.

\begin{deff}
  The \emph{exterior product} is an atisymmetrized version of the tensor product. It is
  defined as:

  \begin{align*}
    \land: \quad \Lambda^*_q\times\Lambda^*_r &\to \Lambda^*_{q+r} \\
    ({}_q\alpha;{}_r\beta) &\mapsto {}_q\alpha\land{}_r\beta \\
    \intertext{such that}
    {}_q\alpha\land{}_r\beta(\vec{v}_1,\dotsc,\vec{v}_{q+r}) \equiv \frac{1}{q!\, r!}\,
      \sum_{\sigma\in\mathcal{S}_{q+r}} (-1)^{|\sigma|}\, & {}_q\alpha(\vec{v}_{\sigma(1)},\dotsc,\vec{v}_{\sigma(q)}) \cdot
      {}_r\beta(\vec{v}_{\sigma(q+1)},\dotsc,\vec{v}_{\sigma(q+r)})
  \end{align*}

  Thus, it is not difficult to see that this product is associative, distributive and
  satisfies: ${}_q\alpha\land{}_r\beta = (-1)^{q\, r}\, {}_r\beta\land{}_q\alpha$. Also, $\dim(\Lambda^*_q) =
  \tfrac{n!}{q!\, (n-q)!}$ (remember that, $n$, labels the dimension of the underlying
  vector space, $\mathds{V}$, while, $q$, labels the degree of the form).
\end{deff}

To make life a bit simpler in the future [when dealing with Clifford Algebras], let's
define the following.

\begin{deff}
  The \emph{interior product}, $i_{\vec v}$, is given by:

  \begin{align*}
    i_{\vec v} : \quad \Lambda^*_q &\to \Lambda^*_{q-1} \\
    {}_q\alpha &\mapsto i_{\vec v}({}_q\alpha) \\
    \intertext{where}
    (i_{\vec v}\, {}_q\alpha)(\vec{\omega}_1,\dotsc,\vec{\omega}_{q-1}) &\equiv
      {}_q\alpha(\vec{v},\vec{\omega}_1,\dotsc,\vec{\omega}_{q-1})
  \end{align*}

  With this, $(i_{\vec v}\, {}_q\alpha)$, is linear in $\vec v$ and ${}_q\alpha$, satisfying:

  \begin{align*}
    i_{\vec v}\, i_{\vec \omega} + i_{\vec \omega}\, i_{\vec v} &= 0 \\
    i_{\vec v}\, ({}_q\alpha\land{}_r\beta) &= (i_{\vec v}\, {}_q\alpha)\land{}_r\beta + (-1)^{q}\, {}_q\alpha\land(i_{\vec v}\, {}_r\beta)
  \end{align*}
\end{deff}

At this point, the last concept to be introduced (before defining the exterior algebra) is
that of a \emph{pull-back} of a $q$-form.

\begin{deff}
  Consider two vector spaces, $\mathds{V}, \mathds{W}$, and $\Lambda^*_q(\mathds{V}),
  \Lambda^*_q(\mathds{W})$ associated with them. Let $f:\; \mathds{V}\to\mathds{W}$, linear. Then,
  it is possible to associate $\forall\; {}_q\alpha\in\Lambda^*_q(\mathds{W})$, a $(f\,
  {}_q\alpha)\in\Lambda^*_q(\mathds{V})$ such that:

  \begin{align*}
    f^*:\quad \Lambda^*_q(\mathds{W)} &\to \Lambda^*_q(\mathds{V}) \\
    {}_q\alpha &\mapsto f\, {}_q\alpha \\
    \intertext{where}
    (f\, {}_q\alpha)(\vec{v}_1,\dotsc,\vec{v}_q) &\equiv {}_q\alpha(f(\vec{v}_1),\dotsc,f(\vec{v}_q))
  \end{align*}
\end{deff}

In this fashion, given that, at our disposal, there is the vector space $(\Lambda^*_q,+,\cdot)$ and
the exterior product, $\land$. However, the set $\Lambda^*_q$ is not closed under $\land$. On the other
hand, $\Lambda^*_q = \{\vec 0\}, \; q > n$ (where $n = \dim(\mathds{V})$). Thus, it is possible to
define:

\begin{deff}
  The structure $(\Lambda^*\equiv\oplus_{q=0}^n\Lambda^*_q,+,\cdot,\land)$ forms the \emph{exterior algebra} over
  $\mathds{V}$. Moreover, $\dim(\Lambda^*) = 2^n$.
\end{deff}

Now, given that $\mathds{V} \simeq \mathds{V}^{**}$ (by virtue of $\vec{v}[\Bvec{x}] \equiv
\Bvec{x}[\vec{v}]$), the very same construction as above can be repeated, \emph{mutatis
  mutandis}, in order to generate de dual of the above algebra.

\begin{deff}
  The structure $(\Lambda\equiv\Lambda^{**}\equiv\oplus_{q=0}^n\Lambda^q,+,\cdot,\land)$ forms the \emph{exterior algebra
  underlying the Clifford algebra} over $\mathds{V}^*$. Moreover, $\dim(\Lambda^) = 2^n$. This
  algebra is also known as the \emph{Clifford exterior algebra}.
\end{deff}
\subsection{Metric and tensors}\label{subsec:mt}
Let's start by restricting $\mathbb{K} = \mathbb{R}$.

\begin{deff}
  A \emph{bilinear} form $g \in \Theta^0_2$ is called a \emph{metric} when it satisfies,

  \begin{align*}
    \text{symmetry:}\quad g(\vec{v},\vec{\omega}) &= g(\vec{\omega},\vec{v}) \\
    \text{regularity:}\quad g(\vec{v},\vec{\omega}) &= 0, \; \forall\vec{\omega}\in\mathds{V} \Rightarrow \vec{v} \equiv
      \vec{0} \; .
    \intertext{By expanding $g$ in a [canonical] basis, the above conditions become:}
    g_{i\, j} &= g_{j\, i} \\
    \det(g_{i\, j}) &\neq 0\; .
  \end{align*}
\end{deff}

The metric further enables the possibility to define a \emph{scalar product} and a
\emph{norm}.

\begin{deff}
  The \emph{scalar product} of two vectors and the \emph{norm} of a vector are defined
  via:

  \begin{align*}
    \vec{v}\bullet\vec{\omega} &= g(\vec{v},\vec{\omega}) = g(\vec{\omega},\vec{v}) = \vec{\omega}\bullet\vec{v} \\
    \norm{v}^2 &= \vec{v}\bullet\vec{v} \; .
  \end{align*}
\end{deff}

The reader should note that nowhere above the restriction that the metric be
\emph{positive definite} was required. Therefore, the metric can, in general, have any
signature. Furthermore, by making use of the metric, the spaces $\mathds{V}$ and
$\mathds{V}^*$ can be related to one another.

In order to accomplish this equivalence, it suffices to say that, for any vector
$\vec{v}\in\mathds{V}$, there is a linear form associated with it, denoted by $\Bvec{v}$,
defined by:

\begin{align*}
  \Bvec{v}:\quad \mathds{V} &\to \mathbb{R} \\
  \vec{x} &\mapsto \Bvec{v}(\vec x) \equiv g(\vec x, \vec v) \; .
\end{align*}

The form $\Bvec{v}$ and the vector $\vec{v}$ are called \emph{metric duals}. This
correspondence can be further pursued and it renders the two exterior algebras,
$(\Lambda^*,+,\cdot,\land)$, $(\Lambda,+,\cdot,\land)$, \emph{isomorphic}, via the metric $g$. Therefore, either of
them can be used in order to construct the Clifford algebra.
\subsection{Clifford algebras and spinors}\label{subsec:cas}
In order to move on to the definition of Clifford Algebras, firstly one needs to define
the \emph{Clifford product}, denoted by $\lor$.

\begin{deff}
  The \emph{Clifford product} is defined via,

  \begin{align*}
    \lor:\quad \Lambda^1\times\Lambda^q &\to \Lambda^{q+1}\oplus\Lambda^{q-1} \\
    (\vec{v},{}^{q}\omega) &\mapsto \vec{v}\lor{}^q\omega \equiv \vec{v}\land{}^q\omega + i_{\vec v}\, {}^q\omega
  \end{align*}

  As it can be seen, this definition makes a heavy use of the isomorphism between
  $\mathds{V} \simeq \mathds{V}^*$.
\end{deff}

A good example is the special case in which $q=1$:

\begin{align*}
  \vec{v}\lor\vec{\omega} &= \vec{v}\land\vec{\omega} + i_{\vec v}(\vec{\omega}) \\
  &= \vec{v}\land\vec{\omega} + g(\vec{v},\vec{\omega}) \\
  \vec{v}\lor\vec{\omega} + \vec{\omega}\lor\vec{v} &= 2\, g(\vec{v},\vec{\omega})
\end{align*}

Also, noting that $\lor$ is associative and distributive, the definition above is extendable
to $\Lambda^r\times\Lambda^q$.

\begin{deff}
  The \emph{Clifford Algebra} can be defined by means of the extension of the definition
  of the Clifford product to the whole of $\Lambda$. Thus, $(\Lambda,+,\cdot,\lor)$ is an algebra over
  $\mathds{V}$. The elements of $\Lambda$ are called \emph{multivectors}.
\end{deff}

Note that, at this level, one has that:

\begin{equation}
  \label{eq:cliff}
  \vec{e}_i\lor\vec{e}_j + \vec{e}_j\lor\vec{e}_i = 2\, g_{i\, j} \; .
\end{equation}

It is not hard to see, then, that the signature of the metric plays an important role in
the Clifford algebra game. Thus, Clifford Algebras are classified by the numbers $r$ and
$s$, which are, respectively, the number of components of the metric which are $+1$ and
$-1$ [note that $r+s = n$, $n=\dim(\mathds{V})$]. Therefore, $(\Lambda,+,\cdot,\lor) \equiv
(\Cliff{r}{s}{R},+,\cdot,\lor)$.

Also, one can make the direct sum decomposition: $\Cliff{r}{s}{R} = \Cliff{r}{s}{R}^+ \oplus
\Cliff{r}{s}{R}^-$, where $\Cliff{r}{s}{R}^+$ is called the \emph{even subalgebra} and
$\Cliff{r}{s}{R}^-$ is called the \emph{odd subalgebra}.

Now, for the sake of completeness (and to satisfy the curiosity of some!), let me state
some results/properties of Clifford Algebras:

\begin{itemize}
\item The \emph{real quaternions}, $\mathds{H}(\mathbb{R})$, are the Clifford Algebra
  \Cliff{0}{2}{R};
\item $\Cliff{3}{1}{R} \simeq \mfr$, where \mfr is the algebra of $4\times4$-square real matrices;
\item $\Cliff{r+1}{s}{R} \simeq \Cliff{s+1}{r}{R}$;
\item $\Cliff{r+1}{s+1}{R} \simeq \Cliff{r}{s}{R}\otimes\Cliff{1}{1}{R}$;
\item $\Cliff{1}{1}{R} \simeq \text{\rsfs M}_{2}(\mathbb{R})$;
\item $\Cliff{1}{3}{R} \simeq \Cliff{0}{2}{R}\otimes\Cliff{1}{1}{R} \simeq
  \mathds{H}(\mathbb{R})\otimes\Cliff{1}{1}{R} \simeq \mathds{H}(\mathbb{R})\otimes\text{\rsfs
  M}_{2}(\mathbb{R})$;
\item $\Cliff{r+1}{s}{R}^+ \simeq \Cliff{s}{r}{R}$;
\item $\Cliff{0}{1}{R} \simeq \mathbb{C}$; \&
\item $\Cliff{3}{1}{R}^+ \simeq \Cliff{1}{2}{R} \simeq \Cliff{0}{1}{R}\otimes\Cliff{1}{1}{R} \simeq
  \Cliff{0}{1}{R}\otimes\text{\rsfs M}_{2}(\mathbb{R}) \simeq \mathbb{C}\otimes\text{\rsfs
  M}_{2}(\mathbb{R}) \simeq \text{\rsfs M}_{2}(\mathbb{C})$.
\end{itemize}
\subsection{Algebraic Definition of \Cliff{r}{s}{R}-spinors}\label{subsec:ads}
The algebraic definition of spinors is very simple, but often obscured in the literature.

Consider a Clifford Algebra $(\Cliff{r}{s}{R},+,\cdot,\lor)$ and its regular representation, $\rho$,
i.e., the mapping from \Cliff{r}{s}{R} to its \emph{endomorphism} [algebraic homomorphism
from a set, $\mathcal{S}$, to itself, denoted by: $\End(\mathcal{S})$] algebra:

\begin{align*}
  \rho: \quad \Cliff{r}{s}{R} &\to \End\big(\Cliff{r}{s}{R}\big) \\
  v &\mapsto \rho_{v}(x) \equiv v\lor x \; .
\end{align*}

In general, $\rho$ is not irreducible, i.e., $\exists \; \mathcal{I} \;\brokenvert \; \mathcal{I} \subset
\Cliff{r}{s}{R}$ which are left invariant under the action of $\rho$: $\rho(\mathcal{I}) \subset
\mathcal{I}$. By definition, such invariant spaces satisfy: $v\lor\mathcal{I} \subset\mathcal{I}$,
i.e., they are \emph{left ideals} of \Cliff{r}{s}{R}. Thus, $\rho$ is \emph{irreducible} when
restricted to any \emph{minimal} left ideal and, then, it is called teh \emph{spinor
  representation}. Different choices of minimal left ideals lead to different but
equivalent representations.
\subsection{Fibre bundles as special manifolds}\label{subsec:fbsm}
\begin{deff}
  A \emph{fibre bundle} $\mathfrak{E}$ over a \cont{\infty} manifold $\mathcal{M}$, with
  typical fibre $\mathcal{F}$ and structure group $\mathcal{G}$, is a set
  $(\mathfrak{E},\mathcal{M,F,G},\pi)$ such that:

  \begin{itemize}
  \item $\mathfrak{E}$ is a \cont{\infty} manifold called the \emph{bundle (or total)
      space};
  \item $\mathcal{M}$ is a \cont{\infty} manifold called the \emph{base space};
  \item $\mathcal{F}$ is a \cont{\infty} manifold called the \emph{typical fibre};
  \item $\mathcal{G}$ is a Lie group acting smoothly on $\mathcal{F}$ on the left, called
    the \emph{structure (or gauge) group};
  \item $\pi$ is a smooth surjection from $\mathfrak{E}$ to $\mathcal{M}$, called the
    \emph{projection}. The set $\pi^{-1}(P), \; P\subset\mathcal{M}$, is called the \emph{fibre over
      $P$}, and is denoted by $\mathcal{F}_P$;
  \item There exists a covering $\{U_i\}$ of $\mathcal{M}$ by open sets, and diffeomorphisms
    $\{\phi_i\}$, mapping $\pi^{-1}(U_i)$ to $U_i\times\mathcal{F}$;
  \item For $Q \in \pi^{-1}(U_i)$, let $\phi_i(Q)$ be written $\phi_i(Q) = (\pi(Q),\psi_{i\, \pi(Q)}(Q)) \in
    U_i\times\mathcal{F}$. Then $\psi_{i\, \pi(Q)}:\; \mathcal{F}_{\pi(Q)} \to \mathcal{F}$ is a
    diffeomorphism; \&
  \item If $P \in U_i \cap U_j$, the mappings $t_{i\, j} \equiv \psi_{i\, P}\circ\psi^{-1}_{j\, P}:\;
  \mathcal{F} \to \mathcal{F}$, are diffeomorphisms called \emph{transition functions} and
  are required to be left-actions by elements of $\mathcal{G}$.
  \end{itemize}
\end{deff}

Loosely speaking, one may say that a fibre bundle $\mathfrak{E}$ over a manifold
$\mathcal{M}$, with typical fibre $\mathcal{F}$ and structure group $\mathcal{G}$, is a
manifold $\mathfrak{E}$ which is locally diffeomorphic to $\mathcal{M}\times\mathcal{F}$, in
such a way that any two points in the fibre $\mathcal{F}_P$ are related smoothly by an
element of $\mathcal{G}$.

Another notion that will be required is that of a \emph{cross-section} of a fibre bundle.

\begin{deff}
  Let $\mathfrak{E}$ be a fibre bundle over $\mathcal{M}$ with typical fibre
  $\mathcal{F}$. By definition/construction, above each point $P \in \mathcal{M}$ there is a
  fibre $\mathcal{F}_P = \pi^{-1}(P)$. A \emph{cross-section}, $\sigma$, of $\mathfrak{E}$ is
  defined as a smooth assignment, for every $P \in \mathcal{M}$, of an element $\sigma_P$ of the
  fibre $\mathcal{F}_P$. Equivalently, it can be defined as being a smooth mapping from
  $\mathcal{M}$ to $\mathfrak{E}$ such that $\pi \circ \sigma$ is the \emph{identity}. Furthermore, a
  cross-section is called \emph{local} iff it is a smooth mapping from an \emph{open}
  subset of $\mathcal{M}$ to $\mathfrak{E}$.
\end{deff}

Among fibre bundles, a special type is more ``commom'' than all the others, the ones
called \emph{principal bundles}.

\begin{deff}
  A \emph{principal bundle} is a fibre bundle in which the typical fibre $\mathcal{F}$ is
  identical with the structure group $\mathcal{G}$.
\end{deff}
\subsection{Geometrical definition of \ctor spinors}\label{subsec:gds}
Consider the manifold $\mathcal{M}$ and is bundle of orthonormal frames with positive
orientation, denoted by $PO^+\mathcal{M}$. The structure group of $PO^+\mathcal{M}$ is
$\mathcal{G} = SO_{3\, 1}$. The transition functions, $t_{i\, j}$, ``acts'' on an
orthonormal frame belonging to a fibre of $PO^+\mathcal{M}$. Moreover, the double-valued
mapping, $\varphi$, from the spin group $\text{Spin}_{r\, s}$ to $SO_{r\, s}$, can be used in
order to construct the set of functions $\tilde{t}_{i\, j}$, elements of $\text{Spin}_{3\,
  1}$, defined by: $\varphi(\tilde{t}_{i\, j}) = t_{i\, j}$.

Thus, given that $t_{i\, j}\, t_{j\, k}\, t_{k\, i} = e$, $t_{i\, i} = e$, and that $\varphi$ is
a \emph{representation} of the Spin group, there are solutions for the above equation for
$\tilde{t}_{i\, j}$ that satisfy: $\tilde{t}_{i\, j}\, \tilde{t}_{j\, k}\, \tilde{t}_{k\,
  i} = \pm e$, $\tilde{t}_{i\, i} = \pm e$. Among them, the \emph{positive} answer will be
chosen. Then, the \emph{principal bundle} $PSF^+\mathcal{M}$ of \emph{spin frames} over
$\mathcal{M}$ is defined as the bundle with $\mathcal{M}$ as the base space, Spin$_{3\,
  1}$ as the structure group and transition functions given by $\tilde{t}_{i\, j}$ chosen
as above.

It should be noted that, because of the 2-to-1 nature of $\varphi$, the bundle
$PSF^+\mathcal{M}$ is a double covering $PO^+\mathcal{M}$.

Furthermore, an assignment of a family of \emph{spin frames} over $\mathcal{M}$ can be
defined as a cross-section of $PSF^+\mathcal{M}$. A spinor above a point $P \in \mathcal{M}$
is a linear combination of the elements of the spin frame given by this cross-section.

In addition to the notion of spinor at a point $P$, one may construct the \emph{spin
  bundle} $\mathfrak{S}\mathcal{M}$ above $\mathcal{M}$. as the bundle which admits as
fibre above the point $P \in \mathcal{M}$ the set of all spinors at $P$. The structure group
of $\mathfrak{S}\mathcal{M}$ is the spin group. A \emph{spinor field} over $\mathcal{M}$
is defined as a cross-section of $\mathfrak{S}\mathcal{M}$.
\subsection{Differentiation of spinor fields}\label{subsec:dsf}
Let's start by noting that the connection 1-forms will be denoted equivalently by
$\mathcal{A}^{\mu}_{\nu}$ and $\gamma^{\mu}_{\nu}$, i.e., $\mathcal{A}^{\mu}_{\nu} = \gamma^{\mu}_{\nu}$. Thus, one
can define a \emph{covariant derivative} for spinor fields in the following way:

\begin{align*}
  \tilde\de_{\vec x} \psi &= \Bigl[\vec{x}(\psi^M) - \tilde{\mathcal{A}}^{M}_{N}(\vec x)\,
    \psi^N\Bigr]\, \tilde{e}_M \\
  &\equiv \vec{x}(\psi^M)\, \tilde{e}_M - \mathcal{D}_{\tilde{\mathcal{A}}}(\psi) \; ,
  \intertext{where}
  \mathcal{D}_{\tilde{\mathcal{A}}}(\psi) &\equiv \tilde{\mathcal{A}}^{M}_{N}(\vec x)\, \psi^N\,
    \tilde{e}_M \; , \\
  \tilde{\mathcal{A}} &= -\frac{1}{2}\, \mathcal{A}_{\mu\, \nu}\, \Sigma^{\mu\, \nu} \; ,\\
  \mathcal{D}_{\tilde{\mathcal{A}}} &= \rho_{\tilde{\mathcal{A}}} \equiv -\frac{1}{2}\,
    \mathcal{A}_{\mu\, \nu}\, \sigma^{\mu\, \nu} \; ,\\
  \intertext{where,}
  4\, \sigma^{\mu\, \nu} &\equiv \gamma^{\mu}\circ\gamma^{\nu} - \gamma^{\nu}\circ\gamma^{\mu} \; ,\\
  &\equiv 4\, \sigma^{\mu\, \nu}{}^{M}_{N}\, \tilde{e}_{M}\otimes\tilde{e}^{N} \; .\\
  \intertext{Therefore, for a \emph{metric compatible} connection, one has that}
  \tilde\de_{\vec x} \psi &= \Bigl[\vec{x}(\psi^M) + \frac{1}{2}\, \mathcal{A}_{\mu\, \nu}(\vec x)\,
    \sigma^{\mu\, \nu}{}^{M}_{N}\, \psi^{N}\Bigr]\, \tilde{e}_M
\end{align*}

Now, in order to define a \emph{Lie derivative} for spinor fields, one could go as
follows: Just make the identification $\mathcal{A}_{\mu\, \nu} = \gamma_{\mu\, \nu} = -L_{\mu\, \nu}$,
where the spinorial Lie derivative is given by $\tilde{\mathcal{L}}_{\vec x}\,
\tilde{e}_{\mu} = -\tilde{e}_{\nu}\, L^{\nu}_{\mu}$. That gives:

\begin{equation*}
  \tilde{\mathcal{L}}_{\vec x}\, \psi = \vec{x}(\psi^M)\, \tilde{e}_M - \frac{1}{2}\, L_{\mu\,
  \nu}\, \sigma^{\mu\, \nu}\, \psi \; .
\end{equation*}
\section{Geometric Calculus and the Ordering Problem in Quantum
  Mechanics}\label{sec:gcopqm}
Given the motivation proposed by \cite{gcroaqtcs01}, one can use the Geometric Calculus
in order to prevent the ordering problem that haunts quantum mechanics. Let's make a quick
overview of the method. Given non-commuting numbers, $\gamma_{\mu}\in \ctor\simeq\mfr$,
(where $\ctor$ is a real Clifford Algebra\footnote{The reader should be aware of the
  adopted notation, where, for a metric $g$, its signature is given by $p$ plus and $q$
  minus signs, with $p+q=n$. The structure of the Clifford Algebra can only depend on $p$
  and $q$, thus it is denoted $\Cliff{p}{q}{K}$, where $\mathbb{K}$ is a field (usually
  taken to be either $\mathbb{R}$ or $\mathbb{C}$). Also, it is a well-known
fact that, $\dim\bigl(\Cliff{p}{q}{K}\bigr)=2^n$.} and $\mfr$ is the space of $4\times 4$
real matrices) it is known that

\begin{equation}
  \label{eq:cliffalg}
  \acomm{\gamma_{\mu}}{\gamma_{\nu}} = \gamma_{\mu}\, \gamma_{\nu} + \gamma_{\nu}\,
    \gamma_{\mu} = 2\, g_{\mu\nu} \; .
\end{equation}
(It should be noted that $\Cliff{2}{0}{R} \simeq \Cliff{1}{1}{R} \simeq
\Matrix{2}{R}$. Thus, one can choose to work either with a 4- or a 2-spinor, depending
only on the choice of the Clifford Algebra desired\footnote{This is a subtle point for
 Quantum Mechanics, given that, once pure spinors are not part of that theory, the type [of
 spinor] that one wants to introduce in the theory is completely arbitrary.}.)

Thus, one can introduce the expansion of an arbitrary vector and the dual basis as
follows\footnote{The reader should note that the common ``slash''-notation is not being
  used here. This should not be a source of [future] confusion.}:
\begin{align*}
  \underline{v} &= \underline{v}^{\mu}\,\gamma_{\mu}\; ; \\
  \acomm{\gamma^{\mu}}{\gamma^{\nu}} &= g^{\mu\nu} \; ; \\
  \gamma^{\mu} &= g^{\mu\nu}\, \gamma_{\nu} \; .
\end{align*}

In the same fashion \cite{cagc84, isg89},
\begin{align*}
  \pa &\equiv \gamma^{\mu}\, \pa_{\mu}\; ; \\
  \pa_{\mu}\, \gamma_{\nu} &= \Gamma^{\alpha}_{\mu\nu}\, \gamma_{\alpha}\; ; \\
  \pa_{\mu}\, \gamma^{\nu} &= -\Gamma^{\nu}_{\mu\alpha}\, \gamma^{\alpha}\; ; \\
  \pa\,\underline{v} &= \gamma^{\mu}\,\gamma^{\nu}\,D_{\mu}\,\underline{v}_{\nu} \; ; \\
  \therefore\quad \pa\,\pa\,\psi &= \gamma^{\mu}\,\gamma^{\nu}\,D_{\mu}\,D_{\nu}\, \psi \;
    ;
\end{align*}
where $\Gamma^{\alpha}_{\mu\nu}$ is the \emph{connection}, $\psi\in\mathbb{K}$ is a
scalar and $D_{\mu} = (\pa_{\mu} - \Gamma^{\alpha}_{\mu\nu})$ is the \emph{covariant}
derivative. If the connection is symmetric (vanishing torsion), then
\begin{equation}
  \label{eq:spinmotion}
  \pa\,\pa\,\psi = D_{\mu}\, D^{\mu}\, \psi \; ;
\end{equation}
which is just D'Alembert's operator in curved spacetime.

Thus, one has that \cite{gcroaqtcs01} ($\hslash = 1$),
\begin{equation}
  \label{eq:popentum}
  \mathfrak{p} \equiv -i\, \pa = -i\, \gamma^{\mu}\, \pa_{\mu} \; ,
\end{equation}
which is Hermitian, $(\ev{\mathfrak{p}} = \langle\hc{\mathfrak{p}}\rangle)$, and whose
expectation value, $(\ev{\mathfrak{p}})$,  follows a \emph{geodesic} trajectory in our
curved spacetime.
\section{Bundle Quantum Mechanics}\label{sec:bqm}
\subsection{Non-Relativistic Quantum Mechanics \cite{fbfqm01}}
The mathematical basis for the reformulation of non-relativistic quantum mechanics in
terms of fibre bundles is given by Schr\"odinger's equation,
\begin{equation*}
  i\, \hslash\, \frac{\dop\psi(t)}{\dop t} = \mathcal{H}(t)\, \psi(t) \; ;
\end{equation*}
$\psi$ is the system's state vector in a suitable Hilbert space $\mathcal{F}$ and
$\mathcal{H}$ is its Hamiltonian.

In the bundle description, one has a Hilbert bundle given by $(F,\pi,\mathcal{M})$, where
the total space is $F$, the projection is $\pi$, the base manifold is $\mathcal{M}$ and a
typical fibre, $\mathcal{F}$,  which is isomorphic to $F_x = \pi^{-1}(x), \; \forall
x\in\mathcal{M}$. Thus, $\exists\; l_x: F_x\to\mathcal{F}, \; x\in\mathcal{M}$,
isomorphisms. A state vector, $(\psi)$, and the Hamiltonian, $(\mathcal{H})$, are
represented respectively by a state section along paths, $\Psi:
\gamma\to\Psi_{\gamma}$, and a bundle Hamiltonian (morphisms along paths)
$\mathcal{H}: \gamma\to\mathcal{H}_{\gamma}$, given by:
\begin{align*}
  \Psi_{\gamma} &: t\mapsto \Psi_{\gamma}(t) = l^{-1}_{\gamma(t)}(\psi(t)) \; ;\\
  \mathcal{H}_{\gamma} &: t\mapsto \mathcal{H}_{\gamma}(t) = l^{-1}_{\gamma(t)}\circ
    \mathcal{H}(t)\circ l_{\gamma(t)} \; ;
\end{align*}
where $\gamma: I\to\mathcal{M}$, $I\subseteq\mathbb{R}$ is the world-line for some
observer. The bundle evolution operator is given by,
\begin{align*}
  U_{\gamma}(t,s) &= l^{-1}_{\gamma(t)}\circ \mathcal{U}(t,s)\circ l^{-1}_{\gamma(s)} :
    F_{\gamma(s)}\to F_{\gamma(t)} \; \\
  \Psi_{\gamma}(t) &= U_{\gamma}(t,s)\, \Psi_{\gamma}(s) \; .
\end{align*}

Therefore, in order to write down the bundle Schr\"odinger equation, $(D\, \Psi = 0)$, a
derivation along paths, $(D)$, corresponding to $U$ is needed,
\begin{equation*}
  D : \text{PLift}^1(F,\pi,\mathcal{M}) \to \text{PLift}^0(F,\pi,\mathcal{M})\; ;
\end{equation*}
where $\text{PLift}^k(F,\pi,\mathcal{M}) = \{\lambda\; \text{lifting} :
\lambda\in\text{\rsfs C\/}^k\}$, is the set of liftings from $\mathcal{M}$ to $F$, and,
\begin{align}
  \nonumber
  \lambda &: \gamma \to \lambda_{\gamma}, \;
    \lambda\in\text{PLift}^1(F,\pi,\mathcal{M}) \; ; \\
  \label{eq:pathd}
  D^{\gamma}_s(\lambda) &= \lim_{\epsilon\to 0}
    \frac{U_{\gamma}(s,s+\epsilon)\lambda_{\gamma}(s+\epsilon) -
    \lambda_{\gamma}(s)}{\epsilon} \; \\
\intertext{where}
  \nonumber
  D^{\gamma}_s(\lambda) &= \bigl[(D\lambda)(\gamma)\bigr](s) = (D\lambda)_{\gamma}(s) \; ;
\intertext{and, in local coords,}
  D^{\gamma}_s(\lambda) &= \biggl(\frac{\dop \lambda_{\gamma}^a(s)}{\dop s} +
    \Gamma^a_b(s;\gamma)\, \lambda^b_{\gamma}(s)\biggr)\, e_a(\gamma(s)) \; ;
\end{align}
where $\{e_a(\gamma(x))\}, s\in I$ is a basis in $F_{\gamma(s)}$. Thus, one can clearly
see what is happening with this structure, namely the bundle evolution transport is giving
origin to the linear connection by means of:
\begin{align*}
  \Gamma_a^b(s;\gamma) &= \frac{\pa\bigl(U_{\gamma}(s,t)\bigr)_a^b}{\pa t}\Bigg\vert_{s=t} =
    -\frac{\pa\bigl(U_{\gamma}(t,s)\bigr)_a^b}{\pa t}\Bigg\vert_{t=s} \; ; \\
  U_{\gamma}(t,s)\, e_a(\gamma(s)) &= \sum_b \bigl(U_{\gamma}(s,t)\bigr)^b_a\,
    e_b(\gamma(t))\; ;
\end{align*}
are the local components of $U_{\gamma}$ in $\{e_a\}$.

In this manner, there is a bijective correspondence between $D$ and the bundle Hamiltonian,
\begin{align*}
  \mathbf{\Gamma}_{\gamma}(t) &= \bigl[\Gamma^b_a(s;\gamma)\bigr] =
    \frac{i}{\hslash}\, \mathbf{H}_{\gamma}(t) \; ; \\
  \mathbf{H}_{\gamma}(t) &= i\, \hslash\, \frac{\pa \mathbf{U}_{\gamma}(t,t_0)}{\pa t}\,
    \mathbf{U}^{-1}_{\gamma}(t,t_0) = \frac{\pa \mathbf{U}_{\gamma}(t,t_0)}{\pa t}\,
    \mathbf{U}_{\gamma}(t_0,t) \; ;
\end{align*}
where $\mathbf{H}_{\gamma}$ is the matrix-bundle Hamiltonian.
\subsection{Relativistic Quantum Mechanics \cite{fbfrqm02}}
\subsubsection{Time-dependent Approach}
The framework developed so far can be generalized to relativistic quantum theories in a
very straightforward way. However, in doing so, it is seen that time plays a privileged
role (thus, leaving the relativistic covariance implicit). Proceeding in such a manner, it
is found that,
\begin{align}
  \label{eq:diracsch}
  i\, \hslash\, \frac{\pa\psi}{\pa t} &= {}^D\!\mathcal{H}\,\psi \; ; \\
  \nonumber
  \psi &= \trans{(\psi_1,\psi_2,\psi_3,\psi_4)} \; ;
\end{align}
where ${}^D\!\mathcal{H}$ is the Dirac Hamiltonian (Hermitian), in the space $\mathcal{F}$
of state spinors $\psi$. Once \eqref{eq:diracsch} is a first-order differential equation,
a \emph{[Dirac] evolution operator}, $({}^D\mathcal{U})$, can be introduced. It is a
$4\times 4$-integral matrix operator uniquely defined by the initial-value problem,
\begin{align*}
  i\, \hslash\, \frac{\pa \psi}{\pa t} {}^D\mathcal{U}(t,t_0) &= {}^D\!\mathcal{H}\circ
    {}^D\mathcal{U}(t,t_0) \\
  {}^D\mathcal{U}(t_0,t_0) &= \mathds{1}_{\mathcal{F}} \; .
\end{align*}

Thus, the formalism developed earlier can be applied to Dirac particles. The spinor
lifting of paths has to be introduced and the \emph{Dirac evolution transport} [along a
path, $\gamma$] is given by,
\begin{equation*}
  {}^D{U}(t,s) = l^{-1}_{\gamma(t)}\circ {}^D\mathcal{U}(t,s)\circ l_{\gamma(s)}, \quad
    s,t\in I \; .
\end{equation*}
The \emph{bundle Dirac equation} is given by (see \eqref{eq:pathd}),
\begin{equation*}
  {}^{D}\!D_{t}^{\gamma}\Psi_{\gamma} = 0 \; .
\end{equation*}
\paragraph{\uwave{Klein-Gordon Equation}}
The spinless, scalar wavefunction $\phi\in\cont{k}, \; k\geqslant 2$, over spacetime,
satisfies the Klein-Gordon equation if (particle of mass $m$, electric charge $e$ and in
an external electromagnetic field given by $A_{\mu}=(\varphi, \vec{A})$),
\begin{equation*}
  \Biggl[\biggl(i\, \hslash\, \frac{\pa}{\pa t} -e\, \varphi\biggr)^2 - c^2\,\biggl(\vec{p} -
    \frac{e}{c}\,\vec{A}\biggr)^2\Biggr]\, \phi = m^2\, c^4\, \phi \; .
\end{equation*}
In order to solve this, a trick can be used. Just let {\tiny $\psi = \trans{\Bigl(\phi +
  \tfrac{i\,\hslash}{m\,c^2}\, \tfrac{\pa\phi}{\pa t}, \phi - \tfrac{i\,\hslash}{m\,c^2}\,
  \tfrac{\pa\phi}{\pa t}\Bigr)}$}. This is a particular good choice if one is interested in
the non-relativistic limit. After some first-order (Schr\"odinger-type) representation of
the Klein-Gordon equation is chosen, the bundle formalism can be applied to spinless
particles.

Thus, the goal is to describe the given equation of motion in terms of some
Schr\"odinger-type operator and then apply the bundle formalism \emph{mutatis mutandis}.
\subsubsection{Covariant Approach}
Now, it will be developed an appropriate covariant bundle description of relativistic
quantum mechanics. The difference between the time-dependent and the covariant formalism
is analogous to the one between the Hamiltonian and the Lagrangian approaches to
relativistic wave equations.
\paragraph{\uwave{Dirac Equation}}
The covariant Dirac equation (for a spin $\tfrac{1}{2}$, mass $m$ and charge $e$ particle,
in an external electromagnetic field $A_{\mu}$) is given by,
\begin{align*}
  \bigl(i\,\hslash\, \Dslash &- m\,c\,\mathds{1}_{4\times 4}\bigr)\, \psi = 0 \; ; \\
  \Dslash &= \gamma^{\mu}\, D_{\mu} \; ; \\
  D_{\mu} &= \pa_{\mu} - \frac{e}{i\, \hslash\, c}\, A_{\mu} \; .
\end{align*}
Since it is a first-order differential equation, it admits an \emph{evolution operator}
$\mathcal{U}$, whose job is to connect different values at different spacetime
points. Thus, for $x_1, x_2 \in M_0$, ($M_0$ being the Minkowski spacetime),
\begin{equation*}
  \psi(x_2) = \mathcal{U}(x_2,x_1)\, \psi(x_1) \; ;
\end{equation*}
where $\mathcal{U}(x_2, x_1)$ is a $4\times 4$-matrix operator, defined as the unique
solution to the initial-value problem:
\begin{align*}
  \bigl(i\,\hslash\, \Dslash - m\, &c\, \mathds{1}_{4\times 4}\bigr)\, \mathcal{U}(x,x_0) =
    0 \; ; \\
  \mathcal{U}(x_0,x_0) &= \mathds{1}_{\mathcal{F}}, \quad x,x_0 \in M_0 \; ;
\end{align*}
where $\mathcal{F}$ is the space of 4-spinors.

Assume that $(F, \pi, \mathcal{M})$ is a vector bundle with total space $F$, projection
$\pi: F\to \mathcal{M}$, fibre $\mathcal{F}$ and isomorphic fibres $F_x =
\pi^{-1}(x), \; x\in \mathcal{M}$. Then, there exists linear isomorphisms $l_x: F_x
\to \mathcal{F}$ --- which are assumed to be \textbf{diffeomorphisms} --- so that
$F_x = l_x^{-1}(\mathcal{F})$ are 4-dim vector spaces.

A \cont{1} section\footnote{$\Psi$ is simply a \emph{section} at this time, as opposed to
  the previous two cases, in which it was a \emph{section along paths}. This corresponds
  to the fact that quantum objects do \emph{not} have trajectories in a classical sense.}
is assigned to a state spinor, $\psi$, i.e., $\Psi \in \text{Sec}^1(F,\pi,\mathcal{M})$,
in the following manner:
\begin{equation*}
  \Psi(x) = l^{-1}_x\bigl(\psi(x)\bigr) \, \in \, F_x = \pi^{-1}(x), \; x\in\mathcal{M} \; .
\end{equation*}
Thus, it follows that:
\begin{align*}
  \Psi(x_2) &= U(x_2,x_1)\, \Psi(x_1) \; , \; x_1,x_2 \in \mathcal{M}\; ; \\
  U(y,x) &= l^{-1}_y\circ \mathcal{U}(y,x)\circ l_x \; : \; F_x \to F_y\; ; \;
    x,y\in\mathcal{M} \; ; \\
  \underbrace{\mathcal{U}(x_3,x_1)}_{\substack{4\times 4 \text{ matrix}\\ \text{operator}} } &=
    \mathcal{U}(x_3,x_2)\circ \mathcal{U}(x_2,x_1) \; ,\; x_1,x_2,x_3 \in \mathcal{M}\; ;
    \\
  \therefore\; U(x_3,x_1) &= U(x_3,x_2) \circ U(x_2,x_1)\; ,\; x_1,x_2,x_3 \in \mathcal{M}
    \; ; \\
  U(x,x) &= \mathds{1}_{F_x} \; , \; x\in\mathcal{M} \; .
\end{align*}
The map $U:\; (y,x)\mapsto U(y,x)$ --- called the \emph{Dirac evolution transport} --- is
a \emph{linear transport along the identity map}, $(\mathds{1}_{\mathcal{M}})$, of
$\mathcal{M}$ in the bundle $(F,\pi,\mathcal{M})$. Thus, it is not difficult to see that
\cite{bp-TP-general, bp-TP-parallelT, bp-TP-morphisms, bp-TM-general},
\begin{align*}
  \mathcal{D}_{\mu}\, \Psi &= 0 \; , \;\; \mu=0,1,2,3 \; ; \\
  \mathcal{D}_{\mu} &= \mathcal{D}_{x^{\mu}}^{\mathds{1}_{\mathcal{M}}} \; .
\end{align*}
These are called \emph{Bundle Dirac Equations}.

At this point, local basis could be introduced and a local view of the above could be
written down. However, from this knowledge, what will be of future interest, will come
from the facts,
\begin{align}
  \nonumber
  G^{\mu} &= l^{-1}_x\circ \gamma^{\mu}\circ l_x \; ; \\
  \label{eq:gelcliff}
  \acomm{G^{\mu}}{G^{\nu}} &= 2\, \eta^{\mu\nu}\, \mathds{1}_{\mathcal{F}} \; ;
\end{align}
where $\eta^{\mu\nu}$ is the Minkowski metric tensor, $[\eta^{\mu\nu}] =
\diag(+1,-1,-1,-1)$. This last equation, \eqref{eq:gelcliff}, is the bundle generalization
of \eqref{eq:cliff}. It is clear that this expression can be put in terms of local
coordinates, in which case it would reduce to (in the equation that follows,
\textbf{boldface} denotes the matrix --- i.e., the expression in local coordinates ---  of
the operator denoted by the same (kernel) symbol),
\begin{equation}
  \label{eq:localcliff}
  \acomm{\mathbf{G}^{\mu}}{\mathbf{G}^{\nu}} = 2\, \eta^{\mu\nu}\, \mathds{1}_{4 \times 4}
    \; ;
\end{equation}
where $\eta^{\mu\nu}$ is the Minkowski metric tensor and $\mathds{1}_{4 \times 4} =
\diag(1,1,1,1)$ is the unit matrix in 4-dim\footnote{The generalization to an arbitrary
  number of dimension is quite clear and straightforward from the equations given.}. Thus,
\eqref{eq:localcliff} is the local expression of \eqref{eq:gelcliff} and a generalization
of \eqref{eq:cliff}.
\section{Putting it all together}\label{sec:piat}
Basically, one wants to generalize eqs \eqref{eq:spinmotion} and \eqref{eq:popentum} to
the bundle formalism previously formulated. In order to do so, let's remember that
($\hslash = 1$, and bringing the ``slash'' notation back):
\begin{align*}
\intertext{\textbf{Bundle} results:}
  G^{\mu} &= l^{-1}_x\circ \gamma^{\mu}\circ l_x \; ; \\
  \mathfrak{d}_{\mu} &= l^{-1}_x\circ \pa_{\mu}\circ l_x \; ; \\
  \dfrakslash &= G^{\mu}(x)\circ \mathfrak{d}_{\mu} \; ; \\
  \dfrakslash &= l^{-1}_x\circ \paslash\circ l_x \; . \\
\intertext{Thus, the bundle version is given by:}
  \mathfrak{p} &= -i\, \dfrakslash = l^{-1}_x\circ (-i\, \paslash)\circ l_x \; ; \\
\intertext{or, in local coords (\textbf{boldface} being matrix-notation, see
\eqref{eq:gelcliff} and \eqref{eq:localcliff}):}
  \mathbf{p} &= -i\, \mathbf{G}^{\mu}\pa_{\mu} \; .
\end{align*}

This works for (non-relativistic) Quan\-tum Me\-cha\-nics and for Relativistic Quan\-tum
Me\-cha\-nics.

When dealing with Quantum Field Theories, somethings have to be said before conclusions
are drawn. Let's start with a quick overview of the relevant facts.
    In the functional [Schr\"odinger's] representation for a \textsc{free scalar} QFT, one
    has that:
    \begin{align*}
      S &= \int \mathcal{L}\, d^4x = \frac{1}{2}\int\bigl(\pa^{\mu}\varphi\pa_{\mu}\varphi -
        m^2\varphi^2\bigr)\, d^4x \; ; \\
    \intertext{where, the conjugate field momentum (note the non-covariant formalism) is:}
      \pi(x) &= \frac{\pa\mathcal{L}}{\pa(\pa_t \varphi)} = \dot{\varphi}(x) \; ; \\
    \intertext{and, the Hamiltonian is:}
      H &= \frac{1}{2}\int\bigl(\pi^2 + \abs{\nabla\varphi}^2 +
      m^2\varphi^2\bigr) \, d^3x \; . \\
    \intertext{The equal-time commutation relations are given by,}
      \comm{\varphi(\vec x,t)}{\pi(\vec y,t)} &= i\delta(\vec{x} - \vec{y}) \; ; \\
      \comm{\varphi(\vec x,t)}{\varphi(\vec y,t)} &= 0 = \comm{\pi(\vec x,t)}{\pi(\vec y,t)}
        \; ; \\
    \intertext{In the coordinate representation, with a basis for the Fock space, where
      $\varphi(\vec x)$ is (now) time independent and diagonal (note that $\phi(\vec x)$ is
      just an ordinary scalar function), one has that:}
      \varphi(\vec x) \ket{\phi} &= \phi(\vec x) \ket{\phi} \; ; \\
      \therefore\; \Psi[\phi] &= \ip{\phi}{\Psi} \; ; \\
      \frac{\delta}{\delta\phi(\vec x)}\phi(\vec y) &= \delta(\vec{x} - \vec{y}) \; \\
      \therefore\; \comm{\frac{\delta}{\delta\phi(\vec x)}}{\phi(\vec{y})} &= \delta(\vec{x} -
        \vec{y})\; . \\
    \intertext{Thus, the functional representation of the equal-time commutators turns out
        to be:}
      \Rightarrow\; \pi(\vec x) &= -i\frac{\delta}{\delta\phi(\vec x)} \; ; \\
      \ipop{\phi'}{\pi(\vec x)}{\phi} &= -i\frac{\delta}{\delta\phi(\vec x)}\delta[\phi' -
        \phi] \; ; \\
    \intertext{and, the momentum operator, $P_i$, which generates spatial displacements, is:}
      \comm{P_j}{\varphi(\vec x,t)} &= -i\frac{\pa}{\pa x^j}\varphi(\vec x,t) \; ; \\
      \therefore\; P_j &= -\int\bigl(\varphi(x)\, \pa_j\pi(x)\bigr)\, d^3x \; ; \\
        P_j &= i\, \int\biggl(\phi(\vec x)\, \pa_j\, \frac{\delta}{\delta\phi(\vec x)}\biggr)\,
        d^3x \; ; \\
    \intertext{thus, using \eqref{eq:popentum}, one has that:}
      \mathfrak{P} &= \gamma^j\, P_j \; . \\
    \intertext{On the other hand, in the momentum representation, $\pi(x)$ is diagonal and time
      independent, which gives:}
      \pi(\vec x) \ket{\varpi} &= \varpi(\vec x)\ket{\varpi} \; ; \\
      \Psi[\varpi] &= \ip{\varpi}{\Psi}\; ; \\
      \varphi(\vec x) &= i\frac{\delta}{\delta\varpi(\vec x)} \; ; \\
        E\, \Psi[\varpi] &= \underbrace{\frac{1}{2}\int\biggl(-\frac{\delta}{\delta\varpi(\vec
        x)}\bigl(-\nabla^2 + m^2\bigr)\frac{\delta}{\delta\varpi(\vec x)} + \varpi^2(\vec
        x)\biggr)\, d^3x}_{= H} \; \Psi[\varpi] \; . \\
    \intertext{Let us introduce a \emph{functional} version of the Fourier transform given by:}
      \Psi[\varpi] &= \int \Psi[\phi]\, e^{i\int\varpi(\vec x)\, \phi(\vec x)\, d^3x} \,
        \mathcal{D}\phi \; . \\
    \intertext{Thus, for a \textsf{free spinor} QFT, analogous relations are valid:}
      H &= \int\hc{\Psi}(x)\bigl( -i\, \gamma^{\mu}\nabla_{\mu} + m\bigr)\Psi(x) \, d^3x \; ; \\
      \acomm{\Psi_{\alpha}(\vec{x},t)}{\hc{\Psi}_{\beta}(\vec{y},t)} &= \delta_{\alpha\beta}\,
        \delta^3(\vec{x} - \vec{y}) \; ; \\
      \acomm{\Psi_{\alpha}(\vec{x},t)}{\Psi_{\beta}(\vec{y},t)} &= 0 =
        \acomm{\hc{\Psi}_{\alpha}(\vec{x},t)}{\hc{\Psi}_{\beta}(\vec{y},t)} \; ; \\
    \intertext{and in the coord\-inate re\-pre\-sen\-ta\-tion,}
      \Psi(\vec x)\ket{\psi} &= \psi(\vec x)\ket{\psi} \; ; \\
    \intertext{where $\Psi(\vec x)$ is an anti\-commuting \textsf{field}, thus $\psi(\vec x)$ must
      be a spin\-or of Grass\-mann func\-tions $\Rightarrow\; \psi^2_{\alpha}(\vec x)=0$, which
        leads us to:}
      \Phi[\psi] &= \ip{\psi}{\Phi} \; ; \\
      \hc{\Psi}_{\beta}(\vec x) &= \frac{\delta}{\delta\psi_{\beta}(\vec x)} \; ; \\
      \therefore\; E\, \Phi[\psi] &= \underbrace{\int\biggl(\frac{\delta}{\delta\psi(\vec
        x)}\, (-i\,\gamma^{\mu}\nabla_{\mu} + m)\, \psi(\vec x)\biggr)\, d^3x}_{= H}\;
        \Phi[\psi] \; ; \\
      \therefore\;\; \mathfrak{E} &= H\, \gamma^0\; .
\end{align*}

Thus, from all of the above, it is not difficult to see that, a four ``spin-vector'' can
be constructed out of:
\begin{align*}
  \mathfrak{E} &= H\, \gamma^0\; ; \\
  \mathfrak{P} &= P_j\, \gamma^j \; ; \\
  \therefore\;\; \mathcal{P} &= (\mathfrak{E},\mathfrak{P}) = \bigl( H\, \gamma^{0} ,
    P_j\, \gamma^{j} \bigr) \; .
\end{align*}
\begin{align*}
\intertext{In a \textbf{co\-vari\-ant} for\-mu\-la\-tion (the con\-served quan\-tity be\-ing the
  ener\-gy-momen\-tum ten\-sor), one would have that:}
  T^{\mu\nu} &= \frac{\pa \mathcal{L}}{\pa\bigl(\pa_{\nu}
    \varphi_a\bigr)}\pa^{\mu}\varphi_a - g^{\mu\nu}\, \mathcal{L} \; . \\
  \therefore\;\; \bigl( T^{\mu\nu}\, \gamma_{\mu}\, \gamma_{\nu} \bigr) &:
    \text{invariant quantity} \; ; \\
\intertext{thus, using the original notation:}
  \mathcal{T}^{\mu\nu} &= l^{-1}_x\circ T^{\mu\nu}\circ l_x \; ; \\
  \mathfrak{T} &= \mathbf{G}_{\mu}\, \mathcal{T}^{\mu\nu}\, \mathbf{G}_{\nu} =
    l^{-1}_x\circ \bigl( T^{\mu\nu}\, \gamma_{\mu}\, \gamma_{\nu} \bigr)\circ l_x \; ;
\end{align*}
The reader should note that, the properties described in \cite{gcroaqtcs01} will easily
generalize to the above cases. This leads us to the following thought.
\begin{conjec}
  The basic description of physical quantities should be done in terms of spin-variables,
  such as spinors, spin-vectors and spin-tensors.
\end{conjec}
\section{Aknowledgements}
This work was partly supported by \textsf{DOE} grant \textsf{DE-FG02-91ER40688 - Task D}\/.
\bibliographystyle{hplain}
\bibliography{bibliography}
\end{document}